\documentclass[aps,prd,reprint]{revtex4-2}

\usepackage{amsmath}
\usepackage{amssymb}
\usepackage{bm}
\usepackage{graphicx}
\usepackage{xcolor} % Comment 2: revision highlighting.
\graphicspath{{./}{C:/Users/DELL/OneDrive/工作池/ALP/CPmix/}}
\usepackage{mathtools}
\usepackage[colorlinks=true,linkcolor=blue,citecolor=blue,urlcolor=blue]{hyperref}
\allowdisplaybreaks[4]

\newcommand{\vect}{\bm}
\newcommand{\epsT}{\epsilon_{\perp}}
\newcommand{\dd}{\mathrm{d}}

\begin{document}

\title{CP Polarimetry with Linearly Polarized Photon Fusion and Double-Tagged Protons}

\author{Qi-Hui Chang}
\email{2024210026@tju.edu.cn}
\affiliation{Department of Physics and Center for Joint Quantum Studies, School of Science,
Tianjin University, Tianjin 300350, China}

\author{Shuai Zhao}
\email{zhaos@tju.edu.cn}
\affiliation{Department of Physics and Center for Joint Quantum Studies, School of Science,
Tianjin University, Tianjin 300350, China}

\date{\today}

\begin{abstract}

Determining the CP phase of a spin-zero resonance coupled to two photons is challenging when its accessible decay modes do not provide a useful CP analyzer. In exclusive photon fusion with both outgoing protons tagged, however, each proton recoil fixes the linear-polarization axis of the corresponding quasi-real photon, providing an event-by-event production-side polarimeter. We evaluate the contraction of the two photon density matrices with the CP-mixed hard scattering tensor and show that scalar--pseudoscalar interference shifts the second harmonic of the signed azimuthal-angle difference between the tagged protons. We then construct normalized cosine and sine moments that allow the CP-mixing angle to be extracted independently of the total production rate and decay kinematics. We quantify the magnetic dilution of the photon polarization and estimate signal yields at 13 and 14 TeV, including elastic soft-survival corrections. Since the same proton tags also reconstruct the central missing mass, the method can in principle be extended to invisibly decaying resonances.

\end{abstract}

\maketitle

\section{Introduction}

The CP property of a neutral spin-zero resonance is
usually inferred from its production and decay.  Its interaction with two photons can involve both $aF_{\mu\nu}F^{\mu\nu}$ and $aF_{\mu\nu}\widetilde F^{\mu\nu}$, allowing scalar and pseudoscalar couplings to contribute to the same exclusive photon-fusion process.  The axionlike particles (ALPs) provide a standard setting for such
interactions~\cite{Peccei:1977hh,Peccei:1977ur,Weinberg:1977ma,Wilczek:1977pj,Kim:1979if,Shifman:1979if,Zhitnitsky:1980tq,Dine:1981rt,Preskill:1982cy,Abbott:1982af,Dine:1982ah,Sikivie:1983ip,Jaeckel:2010ni,Ringwald:2012hr,Irastorza:2018dyq,DiLuzio:2020wdo}, and their collider phenomenology is often described in effective or simplified frameworks~\cite{Mimasu:2014nea,Jaeckel:2015jla,Bauer:2017ris,Brivio:2017ije,Mariotti:2017vtv,Dolan:2017osp,Alonso-Alvarez:2018irt,Gavela:2019wzg,Chala:2020wvs}.  Although the diphoton decay $a\to\gamma\gamma$ provides a clean resonance signature, its two-body kinematics alone cannot determine the CP-mixing phase without the  information of photon polarization. This motivates a production-side polarimeter that determines the polarization state of the incoming photons event by event.

Such a polarimeter can be realized in exclusive photon fusion with both outgoing protons tagged. In $p(p_1)+p(p_2)\to p(p_1')+a(k)+p(p_2')$,  each tagged outgoing proton determines the transverse momentum $\vect q_{rT}$ carried by the emitted photon. Since the dominant elastic photon density is linearly polarized along this direction, the two proton recoils define the polarization axes of photons event by event. The relative orientation of these axes is therefore not only a kinematic variable,  it also carries the interference between scalar and pseudoscalar photon couplings. Scalar–pseudoscalar interference generates a phase shift in the second-harmonic modulation of the signed azimuthal angle difference between the two tagged protons. The tagged-proton azimuth therefore provides a direct production-side analyzer of the CP-mixing phase, without relying on the decay kinematics. For an invisibly decaying resonance, the same proton tags can additionally reconstruct the central missing mass.

The linear polarization of equivalent photons has been studied extensively in ultraperipheral collisions (UPCs).  In coherent
UPCs, the equivalent photon field is linearly
polarized along the transverse electric field~\cite{Bertulani:1987tz,Vidovic:1992ik}.  The
equivalent-photon description of such reactions has a long history and is
widely used in hadronic and nuclear collisions~\cite{Fermi:1924tc,vonWeizsacker:1934nji,Williams:1934ad,Budnev:1974de,Baur:2001jj,Bertulani:2005ru,Baltz:2007kq,Klein:2020fmr}.  Azimuthal modulations arising from linearly polarized photons have been observed in the $\gamma\gamma \to \ell^+\ell^-$ process by the STAR and ATLAS collaborations
\cite{STAR:2019wlg,ATLAS:2020epq,ATLAS:2018pfw,ATLAS:2022yad},  and coherent photon polarization has also been proposed as a probe of more exotic QED bound states such as the leptonium~\cite{Dai:2024imb,dEnterria:2025ecx,Feng:2026yba}.
Heavy ion UPCs are powerful photon-fusion machines because of their
$Z^4$ enhancement and have been used for ALP searches~\cite{Knapen:2016moh,CMS:2018erd,ATLAS:2017fur,ATLAS:2020hii}.  For CP polarimetry, however, they are limited because the two photon transverse momenta are not directly tagged in inclusive ion measurements, and the nuclear coherence scale $1/R_A$ limits the high-mass reach.  Double-tagged $pp$ collisions avoid these limitations by reconstructing both photon transverse momenta event by event, with the accessible mass range set by the proton form factor and the forward-detector $\xi$ acceptance. The substantially higher integrated luminosity available in $pp$ running can also compensate for the absence of the nuclear $Z^4$ enhancement, particularly at high resonance masses~\cite{Shao:2022cly}.

Azimuthal correlations between the outgoing protons have long been studied as probes of the spin-parity and CP properties of centrally produced states.  Kaidalov
\textit{et al.} developed the spin-parity analysis for central exclusive
diffraction, and Khoze, Martin, and Ryskin showed that a signed tagged-proton
azimuthal asymmetry is sensitive to CP-violating Higgs
mixing~\cite{Kaidalov:2003fw,Khoze:2004yb}.  More directly for the channel
considered here, Harland-Lang, Khoze, and Ryskin explicitly showed that the
outgoing-proton azimuthal distribution in exclusive photon fusion
distinguishes scalar from pseudoscalar production and that an asymmetry can
probe potential CP-violating effects~\cite{Harland-Lang:2016qjy}.
The missing-mass method using tagged forward protons was previously explored for an invisibly decaying Higgs boson in central exclusive diffractive production~\cite{Khoze:2004yb}. This strategy was later investigated for more general invisible final states, including those produced via photon–photon fusion, with only the two outgoing protons detected~\cite{Belotsky:2004ex,Khoze:2017igg}.  Those
studies establish the missing-mass search strategy, whereas
Ref.~\cite{Harland-Lang:2016qjy} establishes the CP sensitivity of the
proton azimuth in visible exclusive photon fusion. However, to our
knowledge, these observables have not been combined into a single
measurement strategy for a fully invisible photon-fusion resonance.

In this work, we describe double-tagged exclusive photon fusion using photon density matrices and show that the CP-mixing angle of a spin-zero resonance appears as a phase shift in the second-harmonic modulation of the signed azimuthal-angle difference between the two tagged protons. We construct normalized azimuthal moments that determine this phase independently of the total rate and decay kinematics. The state $a$ can represent either an ALP with explicit
CP-violating couplings or a more general spin-zero extension of the Standard
Model Higgs sector, such as a CP-mixed scalar in a multi-Higgs-doublet
model~\cite{DellAquila:1985mtb,Plehn:2001nj,Choi:2002jk,Buszello:2002uu,Berge:2008wi,Gao:2010qx,Bolognesi:2012mm,Anderson:2013afp,Ellis:2012jv,Freitas:2012kw,Harnik:2013aja}. In this approach, the missing-mass peak identifies the resonance candidate, while the distribution of the signed azimuthal-angle difference between the two tagged protons allows its CP-mixing angle to be extracted.  This production-side, decay-blind CP
spectroscopy is the distinctive capability of this proposal.

The rest of the paper is organized as follows. In Sec.~\ref{sec:densitymatrix},  we establish the photon-polarization framework and derive the CP-sensitive azimuthal structure of the double-tagged process. In Sec.~\ref{sec:observable},  we construct the proton-azimuth observables for extracting the CP-mixing phase. We present the phenomenological analysis in Sec.~\ref{sec:pheno}. A summary of this work is given in Sec.~\ref{sec:summary}.

\section{Photon Polarization and CP-Sensitive Azimuthal Structure}
\label{sec:densitymatrix}

\subsection{Photon Density Matrix}

The elastic photon emission in the double-tagged $pp$ process is described by the standard equivalent-photon formalism~\cite{Drees:1988pp,Kniehl:1990iv}, which also
underlies forward-proton searches for photon-coupled resonances~\cite{Albrow:2008pn,Chapon:2009hh,Fichet:2013gsa,Fichet:2014uka,Fichet:2015vvy,Harland-Lang:2016apc,Baldenegro:2018hng,CMS:2022hly,ATLAS:2023zfc,CMS:2023jgd}. The density-matrix description is particularly useful here because the transverse recoil of each tagged proton retains the linear-polarization information of the emitted photon. We first establish this correspondence before turning to the CP structure of the hard process.

In the equivalent-photon approximation (EPA), the differential cross section for the double-tagged process can be expressed as a contraction of two photon density matrices with the hard-scattering tensor for $\gamma\gamma\to a$, i.e.,
\begin{align}
 \dd\sigma
 \propto
 {\cal H}^{ik,jl}
 \Gamma_1^{ij}(\xi_1,\vect q_{1T})
 \Gamma_2^{kl}(\xi_2,\vect q_{2T})
 \dd\Phi_0 ,
\end{align}
where $i,j,k,l=1,2$ are transverse indices and $\dd\Phi_0$ denotes the
remaining non-angular phase space, flux factors, and cuts.  
The ALP is produced with transverse momentum $\vect k_T=\vect q_{1T}+\vect q_{2T}$, rapidity $1/2 \ln (\xi_1/\xi_2)$, and invariant mass $M_a\simeq \sqrt{\xi_1\xi_2s} $, 
where $\xi_r$($r=1,2$) is the fractional longitudinal momentum loss of proton
$r$.  Thus, the proton tags reconstruct not only the longitudinal photon kinematics through $\xi_1$ and $\xi_2$, but also the transverse directions $\vect q_{1T}$ and $\vect q_{2T}$ that define the two photon-polarization axes event by event.

The object $\Gamma_r^{ij}$ is an elastic electromagnetic photon density matrix with the same transverse-tensor structure as a forward photon transverse-momentum-dependent (TMD) correlator. It is controlled by the elastic proton form factors.
For an unpolarized proton, transverse rotational symmetry leaves only two
leading-power tensor structures.  The correlator can therefore be written
as~\cite{Vidovic:1992ik,Mulders:2000sh,Boer:2011kf}
\begin{align}
 \Gamma_r^{ij}(\xi_r,\vect q_{rT})=& \int \frac{\dd z^- \dd^2 \vect z_T}{(2\pi)^3 \xi_r  p_r^+}e^{-i \xi_r p_r^+ z^-+i \vect q_{rT}\cdot \vect z_T }\nonumber\\
&\times\langle p_r|F^{+i}(z)F^{+j}(0)|p_r\rangle|_{z^+=0} \nonumber\\
 =&
 \frac{\delta^{ij}}{2} f_r
 +
 \left(
 \hat q_{rT}^{i}\hat q_{rT}^{j}
 -\frac{\delta^{ij}}{2}
 \right) h_r,
 \label{eq:photon_density_matrix}
\end{align}
where
$f_r\equiv f_{\gamma/p}(\xi_r,q_{rT}^2)$ is the unpolarized elastic
photon TMD and
$h_r\equiv h_{\gamma/p}(\xi_r,q_{rT}^2)$ is the linearly polarized
elastic photon TMD.  
The
polarization axis is the measured direction of $\vect q_{rT}$, i.e., $\hat{q}_{rT}\equiv \vect{q}_{rT}/|\vect{q}_{rT}|$.  
The direction $\hat{q}$ therefore fixes the polarization axis, while the ratio $\frac{h_{r}}{f_{r}}$ characterizes the degree of linear polarization.

The magnetic contribution adds an azimuthally isotropic component to the photon density matrix, reducing its degree of linear polarization. Using the proton electromagnetic current parametrized by the Dirac and Pauli form factors $F_1$ and $F_2$, and following the elastic equivalent-photon treatment of Refs.~\cite{Drees:1988pp,Kniehl:1990iv}, we obtain $h_r=\mathcal F_{E,r}$ and $f_r=\mathcal F_{E,r}+\mathcal F_{M,r}$, where $q_r=p_r-p_r'$ and $Q_r^2=-q_r^2$, with
\begin{align}
 {\cal F}_{E,r}
 &=
 \frac{\alpha}{\pi^2\xi_r}
 \frac{(1-\xi_r)q_{rT}^2}
 {(q_{rT}^2+\xi_r^2m_p^2)^2}
 F_E(Q_r^2)\,,\nonumber\\
 {\cal F}_{M,r}
 &=
 \frac{\alpha}{\pi^2\xi_r}
 \frac{\xi_r^2}
 {2(q_{rT}^2+\xi_r^2m_p^2)}
 F_M(Q_r^2)\,.
\end{align}
Here $Q_r^2=(q_{rT}^2+\xi_r^2m_p^2)/(1-\xi_r)$
and, with $\tau_r=Q_r^2/(4m_p^2)$,
$G_E=F_1-\tau_rF_2$, $G_M=F_1+F_2$,
$F_E=(G_E^2+\tau_rG_M^2)/(1+\tau_r)=F_1^2+\tau_rF_2^2$, and
$F_M=G_M^2$.
The electric contribution selects linear polarization along $\vect q_{rT}$, while the magnetic contribution is isotropic in transverse polarization space and reduces the degree of linear polarization. Defining the degree of linear polarization as
${\cal P}_{\gamma,r}\equiv h_r/f_r$, one has
\begin{align}
{\cal P}_{\gamma,r}	=1/(1+r_{M,r}), 
\end{align}
with
\begin{align}
	r_{M,r}=\frac{\xi_r^2}{2(1-\xi_r)}
\left(1+\frac{\xi_r^2m_p^2}{q_{rT}^2}\right)\frac{F_M(Q_r^2)}{F_E(Q_r^2)}\, . 
\end{align}

For $q_{rT}^2\gtrsim \xi_r^2m_p^2$, the magnetic dilution is suppressed by $\xi_r^2$, up to the ratio of elastic form factors. The $F_2^2$ contribution to $F_E$ retains the rank-one tensor structure, while only the separate $\mathcal F_{M,r}$ term depolarizes the photon. At small $q_{rT}$, the recoil azimuth becomes poorly resolved, requiring a minimum-recoil cut or an explicit treatment of the detector resolution.

\begin{figure}[ht]
\includegraphics[width=\columnwidth]{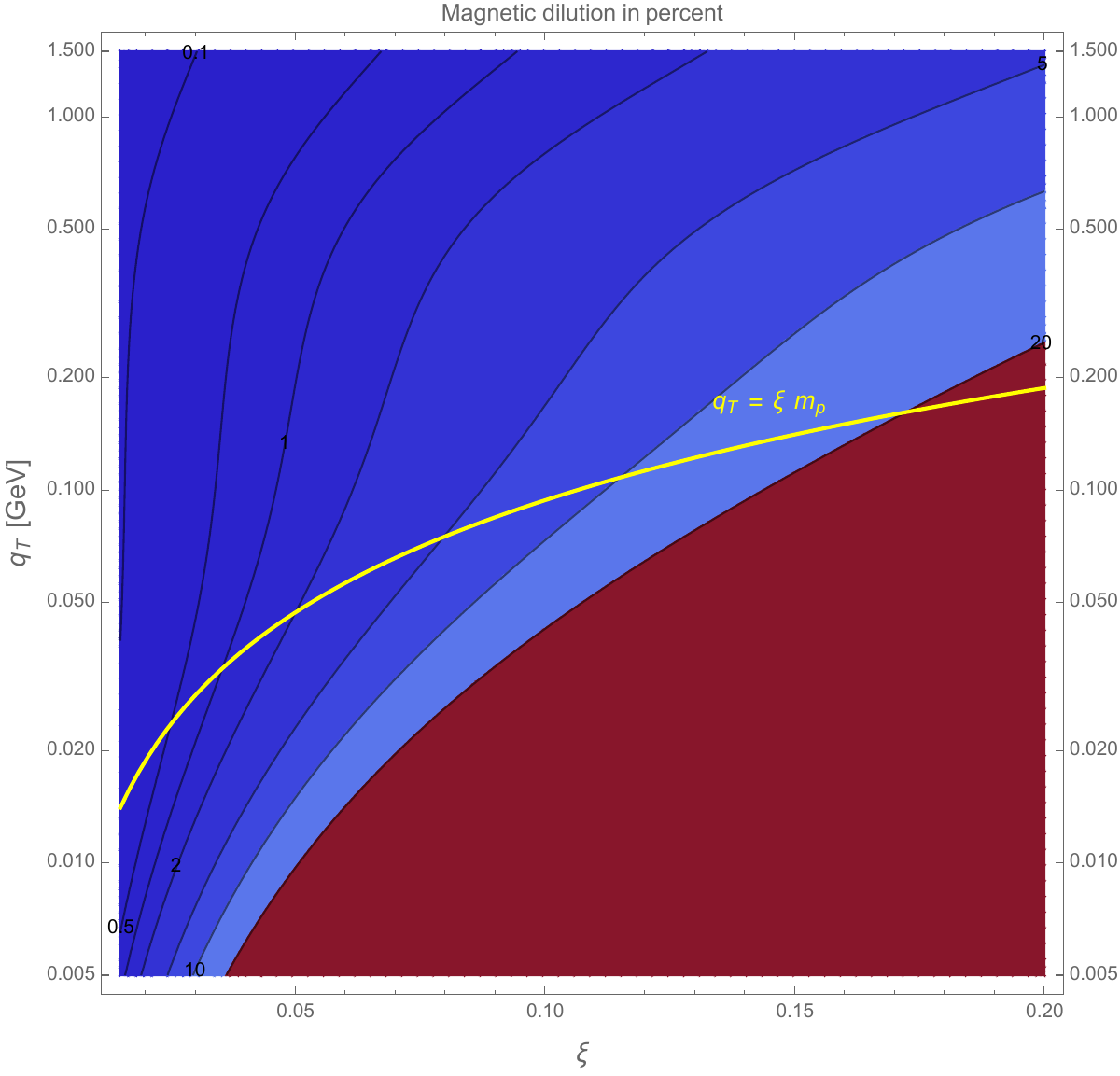}
\caption{
Magnetic dilution $100(1-{\cal P}_\gamma)$ (in percent) of the elastic proton photon
with Sachs dipole form factors.  The yellow curve marks $q_T=\xi m_p$;
away from the high-$\xi$, low-$q_T$ boundary the photon remains nearly
fully linearly polarized.}
\label{fig:proton_polarization_dilution}
\end{figure}

For the numerical illustration in Fig.~\ref{fig:proton_polarization_dilution}
we use the standard dipole form factors in the conventional elastic
photon-flux implementation~\cite{Drees:1988pp,Kniehl:1990iv}.  The $\xi$
range is chosen to cover
AFP/PPS/FP420-like forward-proton acceptances
\cite{Albrow:2008pn,CMS:2022hly,ATLAS:2023zfc,CMS:2023jgd}.
Fig.~\ref{fig:proton_polarization_dilution} shows that sizeable pointwise
dilution is confined to large $\xi$ and $q_T\simeq\xi m_p$, where the
recoil azimuth is also marginal.  Averaging over the elastic photon flux yields $\langle{\cal P}_\gamma\rangle\simeq 0.985$ for $\xi\in[0.015,0.20]$, $q_T\in[0.005,1.5]\,{\rm GeV}$, and $q_T\geq\xi m_p$. For the narrower window $\xi\in[0.015,0.15]$,  $0.05<q_T<0.5\,{\rm GeV}$, and $q_T\geq\xi m_p$, the result is $0.987$. The magnetic contribution therefore causes only a small reduction in the linear polarization, while bins with poorly resolved recoil azimuths require suitable cuts or detector-response modeling.

\subsection{CP-Mixed Hard Amplitude}
\label{sec:hardamp}

Having established the polarization content of the tagged photon emissions, we now turn to the hard $\gamma \gamma\to a$ amplitude and determine how the CP structure is encoded in their relative polarization angle.

We take the two-photon interaction of the spin-zero state to be
\begin{align}
 &{\cal L}_{a\gamma\gamma}
 =
 -\frac{a}{4}
 \left(
 g_S F_{\mu\nu}F^{\mu\nu}
 +g_P F_{\mu\nu}\widetilde F^{\mu\nu}
 \right),
 \label{eq:lagrangian}
\end{align}
where $\widetilde F^{\mu\nu}
=
\frac12\epsilon^{\mu\nu\rho\sigma}F_{\rho\sigma}$ is the dual tensor of
the electromagnetic field. Here $g_S$ and $g_P$ are taken real.
Eq.~\eqref{eq:lagrangian} represents the most general dimension-five
effective interaction of a neutral spin-zero state with two photons.

The transverse amplitude is
obtained by evaluating the $a\gamma\gamma$ vertex between two quasi-real
photons emitted by the tagged protons, with Lagrangian given in Eq.~\eqref{eq:lagrangian}.  For transverse polarizations
$\varepsilon_1^i$ and $\varepsilon_2^k$, the scalar vertex contains
$(q_1\cdot q_2)(\varepsilon_1\cdot\varepsilon_2)$ plus terms suppressed
by $q_r\cdot\varepsilon_s$, while the pseudoscalar vertex contains
$\epsilon_{\mu\nu\rho\sigma}q_1^\mu\varepsilon_1^\nu
q_2^\rho\varepsilon_2^\sigma$.  In the small-angle EPA limit the photons
are transverse and quasi-real, so after removing the common energy factor
$q_1\cdot q_2$ the only remaining two-dimensional tensors are
$\delta^{ik}$ and $\epsT^{ik}$, where
$\epsT^{12}=-\epsT^{21}=+1$.  With the incoming proton 1 chosen along
$+\hat{\vect z}$, this convention fixes the sign of the oriented
azimuthal angle used below.  
The reduced hard amplitude for two transverse photons can be written
as
\begin{align}
 \widehat{\cal M}^{ik}
 =
 g_S\delta^{ik}
 +g_P\epsT^{ik}\,.
\end{align}
The scalar coupling projects parallel linear polarizations, while the
pseudoscalar coupling projects perpendicular linear polarizations.  

The
corresponding hard tensor is
\begin{align}
 {\cal H}_{SP}^{ik,jl}
 =&\,
 g_S^2\delta^{ik}\delta^{jl}
 +g_P^2\epsT^{ik}\epsT^{jl}
 +
 g_Sg_P
 \left(
 \delta^{ik}\epsT^{jl}
 +\epsT^{ik}\delta^{jl}
 \right).
 \label{eq:hard_tensor}
\end{align}
The first two terms are CP even, while the last term is the
scalar--pseudoscalar interference.  It is this $g_S g_P$ interference term that generates the CP-sensitive sine harmonic below. The differential cross section is therefore
\begin{align}
 \dd\sigma
 \propto
 {\cal H}_{SP}^{ik,jl}
 \Gamma_1^{ij}\Gamma_2^{kl}\,\dd\Phi_0 ,
\end{align}
where $d\Phi_0$ denotes the remaining phase-space measure and kinematic factors, with the relevant cuts understood.

To make the azimuthal structure explicit, we define the oriented relative angle between the two photon transverse momenta as
\begin{align}
	\varphi
	=
	\operatorname{atan2}
	(
	\epsT^{ij}q_{2T}^{i}q_{1T}^{j},
	\vect q_{1T}\cdot\vect q_{2T}
	).
\end{align}

With $\epsT^{12}=+1$, this definition gives
$\varphi=\phi_{q_1}-\phi_{q_2}$.  Hence
$\hat{\vect q}_{1T}\cdot\hat{\vect q}_{2T}=\cos\varphi$, while
$\epsT^{ik}\hat q_{1T}^i\hat q_{2T}^k=-\sin\varphi$. Choosing incoming proton 1 along $+\hat{\vect z}$ fixes the orientation convention for $\phi$. Reversing the definition of the signed angle reverses the sine term, but leaves the physical phase shift unchanged.

Introducing the mixing angle $\chi$ so that $g_S=g\cos\chi$, and $g_P=g\sin\chi$, then with the density matrix in Eq.~(\ref{eq:photon_density_matrix}), the hard
contraction becomes
\begin{align}
	{\cal H}_{SP}^{ik,jl}
	\Gamma_1^{ij}\Gamma_2^{kl}
	=
	\frac{g^2}{2}f_1f_2
	\left[
	1+{\cal P}_{\gamma,1}{\cal P}_{\gamma,2}
	\cos(2\varphi+2\chi)
	\right], 
	\label{eq:phase_shift_local}
\end{align}
For photons linearly polarized along their respective transverse momenta, the hard amplitude is proportional to $g_S\cos\varphi-g_P\sin\varphi=g\cos(\varphi+\chi)$, yielding the $\cos(2\varphi+2\chi)$ dependence in Eq.~(\ref{eq:phase_shift_local}).
The photon polarization determines the magnitude of the second-harmonic modulation, while the CP-mixing angle determines its phase.

\section{Tagged-Proton Azimuthal Observables}
\label{sec:observable}

The relative azimuthal angle $\varphi$ of the two photon transverse momenta can be reconstructed from the tagged-proton recoils. Defining $\phi_r=\arg(p'_{rx}+ip'_{ry})$ and using $\vect q_{rT}=-\vect p'_{rT}$, we obtain
\begin{align}
\Delta\phi_{pp}\equiv\phi_1-\phi_2=\varphi
\quad(\mathrm{mod}\ 2\pi),
\end{align}
since reversing both transverse momenta leaves their azimuthal difference unchanged.
Equivalently,
$\Delta\phi_{pp}=\operatorname{atan2}[
\hat{\vect z}\cdot(\vect p_{2T}'\times\vect p_{1T}'),
\vect p_{1T}'\cdot\vect p_{2T}']$, with the convention chosen above.  The signed azimuthal-angle difference preserves the sine harmonic arising from scalar–pseudoscalar interference, which vanishes when the distribution is folded under $\Delta\phi_{pp}\to-\Delta\phi_{pp}$.
After integration over a kinematic bin ${\cal B}$, the distribution keeps
the same form,
\begin{align}
 \frac{\dd\sigma_{\cal B}}{\dd\Delta\phi_{pp}}
 =
 {\cal N}_{\cal B}
 \left[
 1+
 {\cal P}_{\gamma\gamma}({\cal B})
 \cos(2\Delta\phi_{pp}+2\chi)
 \right],
 \label{eq:bin_distribution}
\end{align}
where, before absorptive and detector corrections,
\begin{align}
 {\cal P}_{\gamma\gamma}({\cal B})
 =
 \frac{
 \displaystyle\int_{\cal B}\dd\Phi_0\,h_1h_2
 }{
 \displaystyle\int_{\cal B}\dd\Phi_0\,f_1f_2
 } .
 \label{eq:effective_polarization}
\end{align}
Physically, the tagged-proton recoils determine the linear-polarization axes of the two equivalent photons event by event. The CP-mixing angle is therefore encoded as a phase shift of the second harmonic in the signed azimuth $\Delta\phi_{pp}$.

\begin{figure}[htbp]
	\includegraphics[width=\columnwidth]{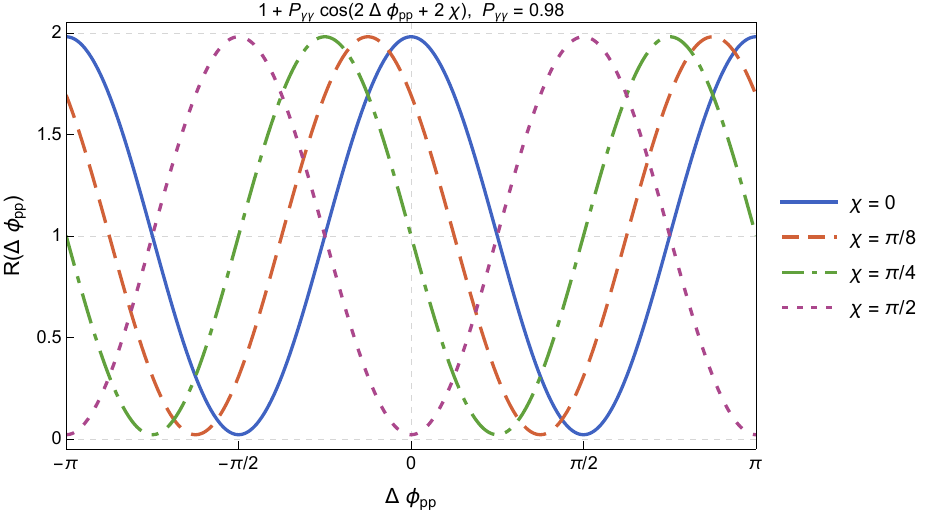}
	\caption{
		Direct visualization of the primary prediction in
		Eq.~(\ref{eq:bin_distribution}).  The plotted quantity is
		$R(\Delta\phi_{pp})\equiv
		1+{\cal P}_{\gamma\gamma}\cos(2\Delta\phi_{pp}+2\chi)$, with the
		representative value ${\cal P}_{\gamma\gamma}=0.98$. The CP phase enters the harmonic argument as $+2\chi$, corresponding to a displacement of its maxima to $\Delta\phi_{pp}=-\chi\pmod{\pi}$ under our signed-azimuth convention.}
	\label{fig:phase_shift_distribution}
\end{figure}

Fig.~\ref{fig:phase_shift_distribution} shows the production-side
polarimeter before any projection onto an asymmetry.  A pure scalar peaks
when the two photon-polarization axes are parallel, while a pure
pseudoscalar peaks when they are perpendicular.  A CP-mixed state translates
the full second harmonic.  This makes the CP phase directly visible as an angular displacement of the modulation.

To extract this phase without relying on the overall normalization, we introduce the normalized second moments  
\begin{align}
 {\cal C}_2^{pp}({\cal B})
 &\equiv
 2\left\langle\cos2\Delta\phi_{pp}\right\rangle_{\cal B}
 =
 {\cal P}_{\gamma\gamma}({\cal B})\cos2\chi,
 \nonumber\\
 {\cal S}_2^{pp}({\cal B})
 &\equiv
 2\left\langle\sin2\Delta\phi_{pp}\right\rangle_{\cal B}
 =
 -{\cal P}_{\gamma\gamma}({\cal B})\sin2\chi .
 \label{eq:moments}
\end{align}
The CP-mixing angle and the modulation amplitude can then be extracted as
\begin{align}
	\chi
	=
	\frac{ \operatorname{atan2}
		\left(-{\cal S}_2^{pp},{\cal C}_2^{pp}\right)}{2},
	~~{\cal P}_{\gamma\gamma}
	=
	\sqrt{({\cal C}_2^{pp})^2+({\cal S}_2^{pp})^2}.
	\label{eq:phase_extraction}
\end{align}
Eq.~\eqref{eq:phase_extraction} is the Born-level estimator. Soft screening dilutes the cosine and sine moments differently, so extracting $\chi$ from screened events requires the corresponding corrections to both moments.

At Born level, the same sine modulation gives the counting asymmetry,
\begin{align}
	A_{\rm CP}^{pp}({\cal B})
	&=
	\frac{
		N_{\cal B}(\sin 2\Delta\phi_{pp}>0)
		-N_{\cal B}(\sin 2\Delta\phi_{pp}<0)}
	{N_{\cal B}(\sin 2\Delta\phi_{pp}>0)
		+N_{\cal B}(\sin 2\Delta\phi_{pp}<0)}
\nonumber	\\
	&=\frac{2}{\pi}{\cal S}_2^{pp}.
\end{align}

\section{Phenomenological Analysis}
\label{sec:pheno}

We now evaluate the azimuthal modulation and expected signal yields under representative LHC conditions. As a benchmark, we take $\sqrt{s}=13~{\rm TeV}$ and a forward-proton acceptance of $0.02<\xi<0.1$.  This choice fixes the accessible double-tagged mass
window to
\begin{align}
 \xi_{\min}\sqrt{s}<M_X<\xi_{\max}\sqrt{s},
 ~~ M_X\simeq260\text{--}1300\,{\rm GeV}.
\end{align}
Thus the standard AFP/PPS-like double-tag configuration is naturally a
high-mass polarimeter. 
Benchmarks with smaller masses such as $M_a=10$ or $100\,{\rm GeV}$ lie outside the assumed acceptance.  This limitation is purely kinematic and does not weaken
the CP observable in the mass window that can actually be tagged.

For a narrow resonance of fixed mass, the longitudinal momentum fractions
can be parameterized as
\begin{align}
 \xi_{1,2}(y)
 &=
 \frac{M_a}{\sqrt{s}}e^{\pm y},
 \qquad \xi_1\xi_2s\simeq M_a^2 .
\end{align}
After integration over the transverse recoils, define
\begin{align}
 I_f(\xi)
 &=
 \int_{q_T^{\min}(\xi)}^{q_T^{\max}}
 q_T\dd q_T\,f_{\gamma/p}(\xi,q_T^2),
 \nonumber\\
 I_h(\xi)
 &=
 \int_{q_T^{\min}(\xi)}^{q_T^{\max}}
 q_T\dd q_T\,h_{\gamma/p}(\xi,q_T^2).
\end{align}
Eq.~(\ref{eq:effective_polarization}) then becomes
\begin{align}
 {\cal P}_{\gamma\gamma}(M_a)
 =
 \frac{
 \displaystyle\int_{y_{\min}}^{y_{\max}}\dd y\,
 I_h(\xi_1)I_h(\xi_2)
 }{
 \displaystyle\int_{y_{\min}}^{y_{\max}}\dd y\,
 I_f(\xi_1)I_f(\xi_2)
 }.
\end{align}
Here the rapidity range is fixed by $0.02<\xi_{1,2}<0.1$, and we use
the
dipole form factors specified above. We restrict the numerical illustration to the transverse-recoil-dominated
region
\begin{align}
	\xi m_p \leq q_T \leq 1.5\,{\rm GeV}.
\end{align}
The auxiliary $5\,{\rm MeV}$ lower cutoff used in the numerical
implementation is inactive for $0.02<\xi<0.1$.
For $M_a=300,500,800$, and $1000\,{\rm GeV}$,
we obtain
\begin{align}
	{\cal P}_{\gamma\gamma}
	&\simeq (0.996,0.984,0.964,0.949),
	\nonumber\\
	|A_{\rm CP}^{pp}|_{\max}
	&=\frac{2}{\pi}{\cal P}_{\gamma\gamma}
	\simeq (0.63,0.63,0.61,0.60),
\end{align}
respectively, where the maximal counting asymmetry $|A_{\rm CP}^{pp}|$ is
reached near $\chi=\pi/4$.
The mass dependence follows from the single photon polarization
${\cal P}_{\gamma,r}=1/(1+r_{M,r})$.  In the resolved-recoil region,
$r_{M,r}$ scales parametrically as $\xi_r^2F_M/F_E$.  A larger $M_a$
therefore selects larger typical $\xi_r$ and enhances the isotropic
magnetic contribution, producing the mild decrease of
${\cal P}_{\gamma\gamma}$.  Nevertheless, the second-harmonic modulation
remains close to its fully polarized limit throughout the tagged mass
window. The experimental sensitivity to the CP-mixing angle therefore depends on the signal yield after soft-survival suppression and the efficiency of the event selection.

For the Born distribution in Eq.~\eqref{eq:bin_distribution} and $N$ independent signal events, the exact leading statistical covariance matrix is
$
\operatorname{Var}({\cal C}_2^{pp})=\frac{2-({\cal C}_2^{pp})^2}{N},
\operatorname{Var}({\cal S}_2^{pp})=\frac{2-({\cal S}_2^{pp})^2}{N},
\operatorname{Cov}({\cal C}_2^{pp},{\cal S}_2^{pp})
=-\frac{{\cal C}_2^{pp}{\cal S}_2^{pp}}{N}
$. Propagating this covariance matrix through the phase estimator gives
$
\delta\chi=
\frac{1}{\sqrt{2N}\,{\cal P}_{\gamma\gamma}}
$, independently of $\chi$. The statistical uncertainty is therefore independent of the CP-mixing angle.

We next estimate the signal yields that determine the
statistical precision. In the narrow-width approximation,
the  Born-level exclusive production cross section is given by
\begin{align}
 \sigma_{\rm Born}(pp\to p+a+p)
	&=
	\frac{8\pi^2}{M_a}\,
	\Gamma_{a\to\gamma\gamma}
	\left.
	\frac{d{\cal L}_{\gamma\gamma}}{dM^2}
	\right|_{M_a},
	\nonumber\\
	\Gamma_{a\to\gamma\gamma}
	&=
	\frac{g_{a\gamma\gamma}^2M_a^3}{64\pi},
	\label{eq:rate_estimate}
\end{align}
where $g_{a\gamma\gamma}^2=g_S^2+g_P^2$.

For the numerical estimates, we adopt
$g_{a\gamma\gamma}=0.3\,{\rm TeV}^{-1}$
as a reference benchmark.
For ${\rm Br}(a\to\gamma\gamma)=1$,
ATLAS reports upper limits on $g_{a\gamma\gamma}$
of $0.04$--$0.09\,{\rm TeV}^{-1}$
over the mass range $150$--$1600\,{\rm GeV}$,
with comparable constraints from CMS--TOTEM PPS.
The reference coupling is therefore excluded for an
unsuppressed diphoton branching fraction.
 In the narrow-width approximation, production and decay factorize, so the event yield for $a\to f$ scales as $g_{a\gamma\gamma}^2{\rm Br}(a\to f)$. The branching fraction cancels in the normalized azimuthal moments.
 Using the same elastic photon
densities and $0.02<\xi<0.1$, we find
\begin{align}
\sigma_{\rm Born}(pp\to p+a+p)
\simeq
0.45,1.19,0.48,0.19~{\rm fb}
\end{align}
for $M_a=300,500,800,$ and $1000\,{\rm GeV}$, respectively. For the event-yield estimates at $\chi=\pi/4$, we use the screened cross sections computed with a modified native SuperChic~4.2~\cite{Harland-Lang:2020veo} implementation of the on-shell EPA and KMR model~4~\cite{Khoze:2013dha} for the stated proton cuts at each collision energy. The survival probabilities vary by approximately $0.2\%$–$0.6\%$ among KMR models 1–4 for the 13 TeV benchmarks. The predicted rates also depend on the elastic proton form factors, while the event yields scale with the channel-dependent reconstruction and selection efficiencies.

For the reference yield estimates, we assume an illustrative efficiency $\epsilon_{\rm tag}\epsilon_{\rm excl}=0.5$. Here $\epsilon_{\rm tag}$ denotes the efficiency for reconstructing both forward protons within the specified acceptance, while $\epsilon_{\rm excl}$ accounts for central-system selection, exclusivity requirements, and kinematic matching between the tagged protons and the reconstructed central system, including timing constraints where applicable.
Neither efficiency includes the soft-survival probability $S^2$ or the
branching fraction.   Using the screened cross section, which already includes soft-survival suppression, the signal yield is
\begin{align}
N_{\rm sig}={\cal L}_{\rm int}\,\sigma_{\rm scr}\,{\rm Br}(a\to f)\,\epsilon_{\rm tag}\epsilon_{\rm excl}.
\end{align}
Table~\ref{tab:comment2_yields} gives the reference yields obtained from these screened cross sections at $13$ and $14\,{\rm TeV}$ for ${\cal L}_{\rm int}=300$ and $3000\,{\rm fb}^{-1}$. At $M_a=300\,{\rm GeV}$, the rate decreases at higher energy because the accepted rapidity interval narrows near $M_{\min}=0.02\sqrt{s}$.
The yield entries scale as
\begin{align}
\left(\frac{g_{a\gamma\gamma}}{0.3\,{\rm TeV}^{-1}}\right)^2
\left[
 \frac{{\rm Br}(a\to f)\,\epsilon_{\rm tag}\epsilon_{\rm excl}}
{0.5}
\right].
\end{align}
\begin{table}[!htb]
\caption{Screened cross sections and reference signal yields for $g_{a\gamma\gamma}=0.3\,{\rm TeV}^{-1}$, $\chi=\pi/4$, ${\rm Br}(a\to f)=1$, and $\epsilon_{\rm tag}\epsilon_{\rm excl}=0.5$, with the stated proton cuts. The yields $N_{300}$ and $N_{3000}$ correspond to integrated luminosities of $300$ and $3000\,{\rm fb}^{-1}$, respectively.}
\label{tab:comment2_yields}
\centering
\footnotesize
\setlength{\tabcolsep}{3pt}
\begin{tabular}{ccr rr}
\hline\hline
$\sqrt{s}$ [TeV] & $M_a$ [GeV] & $\sigma_{\rm scr}$ [fb] & $N_{300}$ & $N_{3000}$ \\
\hline
13 & 300 & $0.367$ & 55.0 & 550.3 \\
13 & 500 & $0.876$ & 131.4 & 1313.8 \\
13 & 800 & $0.309$ & 46.4 & 463.9 \\
13 & 1000 & $0.109$ & 16.3 & 163.3 \\
\hline
14 & 300 & $0.191$ & 28.6 & 285.9 \\
14 & 500 & $0.865$ & 129.8 & 1297.5 \\
14 & 800 & $0.403$ & 60.4 & 604.2 \\
14 & 1000 & $0.162$ & 24.3 & 242.5 \\
\hline\hline
\end{tabular}
\end{table}

For the visible channel $a\to\gamma\gamma$, the main backgrounds are the exclusive diphoton continuum and inclusive or proton-dissociative diphoton production accompanied by accidental proton tags. Matching the mass and rapidity reconstructed from the tagged protons to those of the diphoton system, together with proton timing, suppresses accidental associations~\cite{TOTEM:2023ewz,Cerny:2020rvp}. Mass sidebands and event mixing can constrain the residual background yields and azimuthal shapes, including the second harmonics entering the phase extraction.

The method can also be applied to invisibly decaying resonances. In exclusive production, the two tagged protons reconstruct the missing four-momentum and invariant mass,
\begin{align}
M_{\rm miss}^2=(p_1+p_2-p_1'-p_2')^2\simeq \xi_1\xi_2s.
\end{align}
A localized excess in the missing-mass spectrum would indicate a candidate resonance. The azimuthal distribution of the same tagged protons could then be used to extract its CP-mixing angle without observing the decay products.

For searches for invisibly decaying resonances, unrelated forward protons from pileup can be mistaken for protons from exclusive production. Single- and double-dissociative photon-fusion processes may also enter the double-tagged sample when the dissociation products escape detection and the missing proton tags arise from unrelated interactions. Proton timing, exclusivity requirements, and low-pileup running can help reduce these backgrounds~\cite{Belotsky:2004ex,Khoze:2017igg,Harland-Lang:2018hmi,Goncalves:2020saa,Martins:2022dfg,Tasevsky:2022sch,Cerny:2020rvp}.

\section{Summary}
\label{sec:summary}

We have shown that the transverse recoils of the two tagged protons in exclusive photon fusion determine the linear-polarization axes of the incoming photons. Scalar–pseudoscalar interference produces a phase shift in the second harmonic of the signed azimuthal-angle difference between the tagged protons. The corresponding normalized cosine and sine moments allow the CP-mixing angle of a spin-zero resonance to be extracted after correcting for soft rescattering, without using its decay products as CP analyzers. For representative AFP/PPS acceptance, the Born-level second-harmonic modulation remains close to its fully polarized limit, with only a small magnetic dilution.

The same proton tags also reconstruct the central missing mass, allowing the method to be extended in principle to invisibly decaying resonances.  The sensitivity to the CP-mixing angle depends on the signal yield, soft-survival effects, selection efficiencies, and backgrounds, particularly accidental forward-proton tags at high pileup. A dedicated detector-level study is needed to determine the achievable sensitivity.

\begin{acknowledgments}
The authors thank W.~Z.~Feng for useful discussions and M.~Tasevsky for correspondence.
This work was supported in part by the National Natural Science Foundation
of China under Contract No.~12475098.
\end{acknowledgments}


\begin{thebibliography}{99}
	
	%\cite{Peccei:1977hh}
	\bibitem{Peccei:1977hh}
	R.~D.~Peccei and H.~R.~Quinn,
	%``CP Conservation in the Presence of Instantons,''
	Phys. Rev. Lett. \textbf{38}, 1440-1443 (1977)
	doi:10.1103/PhysRevLett.38.1440
	%9734 citations counted in INSPIRE as of 22 Oct 2026
	
	%\cite{Peccei:1977ur}
	\bibitem{Peccei:1977ur}
	R.~D.~Peccei and H.~R.~Quinn,
	%``Constraints Imposed by CP Conservation in the Presence of Instantons,''
	Phys. Rev. D \textbf{16}, 1791-1797 (1977)
	doi:10.1103/PhysRevD.16.1791
	%5041 citations counted in INSPIRE as of 22 Oct 2026
	
	%\cite{Weinberg:1977ma}
	\bibitem{Weinberg:1977ma}
	S.~Weinberg,
	%``A New Light Boson?,''
	Phys. Rev. Lett. \textbf{40}, 223-226 (1978)
	doi:10.1103/PhysRevLett.40.223
	%7044 citations counted in INSPIRE as of 22 Oct 2026
	
	%\cite{Wilczek:1977pj}
	\bibitem{Wilczek:1977pj}
	F.~Wilczek,
	%``Problem of Strong  $P$  and  $T$  Invariance in the Presence of Instantons,''
	Phys. Rev. Lett. \textbf{40}, 279-282 (1978)
	doi:10.1103/PhysRevLett.40.279
	%6760 citations counted in INSPIRE as of 22 Oct 2026
	
	%\cite{Kim:1979if}
	\bibitem{Kim:1979if}
	J.~E.~Kim,
	%``Weak Interaction Singlet and Strong CP Invariance,''
	Phys. Rev. Lett. \textbf{43}, 103 (1979)
	doi:10.1103/PhysRevLett.43.103
	%3717 citations counted in INSPIRE as of 21 Oct 2026
	
	%\cite{Shifman:1979if}
	\bibitem{Shifman:1979if}
	M.~A.~Shifman, A.~I.~Vainshtein and V.~I.~Zakharov,
	%``Can Confinement Ensure Natural CP Invariance of Strong Interactions?,''
	Nucl. Phys. B \textbf{166}, 493-506 (1980)
	doi:10.1016/0550-3213(80)90209-6
	%3378 citations counted in INSPIRE as of 21 Oct 2026
	
	%\cite{Zhitnitsky:1980tq}
	\bibitem{Zhitnitsky:1980tq}
	A.~R.~Zhitnitsky,
	%``On Possible Suppression of the Axion Hadron Interactions. (In Russian),''
	Sov. J. Nucl. Phys. \textbf{31}, 260 (1980)
	%2761 citations counted in INSPIRE as of 21 Oct 2026
	
	%\cite{Dine:1981rt}
	\bibitem{Dine:1981rt}
	M.~Dine, W.~Fischler and M.~Srednicki,
	%``A Simple Solution to the Strong CP Problem with a Harmless Axion,''
	Phys. Lett. B \textbf{104}, 199-202 (1981)
	doi:10.1016/0370-2693(81)90590-6
	%4157 citations counted in INSPIRE as of 21 Oct 2026
	
	%\cite{Preskill:1982cy}
	\bibitem{Preskill:1982cy}
	J.~Preskill, M.~B.~Wise and F.~Wilczek,
	%``Cosmology of the Invisible Axion,''
	Phys. Lett. B \textbf{120}, 127-132 (1983)
	doi:10.1016/0370-2693(83)90637-8
	%4438 citations counted in INSPIRE as of 22 Oct 2026
	
	%\cite{Abbott:1982af}
	\bibitem{Abbott:1982af}
	L.~F.~Abbott and P.~Sikivie,
	%``A Cosmological Bound on the Invisible Axion,''
	Phys. Lett. B \textbf{120}, 133-136 (1983)
	doi:10.1016/0370-2693(83)90638-X
	%4064 citations counted in INSPIRE as of 22 Oct 2026
	
	%\cite{Dine:1982ah}
	\bibitem{Dine:1982ah}
	M.~Dine and W.~Fischler,
	%``The Not So Harmless Axion,''
	Phys. Lett. B \textbf{120}, 137-141 (1983)
	doi:10.1016/0370-2693(83)90639-1
	%4031 citations counted in INSPIRE as of 22 Oct 2026
	
	%\cite{Sikivie:1983ip}
	\bibitem{Sikivie:1983ip}
	P.~Sikivie,
	%``Experimental Tests of the Invisible Axion,''
	Phys. Rev. Lett. \textbf{51}, 1415-1417 (1983)
	[erratum: Phys. Rev. Lett. \textbf{52}, 695 (1984)]
	doi:10.1103/PhysRevLett.51.1415
	%2413 citations counted in INSPIRE as of 21 Oct 2026
	
	%\cite{Jaeckel:2010ni}
	\bibitem{Jaeckel:2010ni}
	J.~Jaeckel and A.~Ringwald,
	%``The Low-Energy Frontier of Particle Physics,''
	Ann. Rev. Nucl. Part. Sci. \textbf{60}, 405-437 (2010)
	doi:10.1146/annurev.nucl.012809.104433
	[arXiv:1002.0329 [hep-ph]].
	%1277 citations counted in INSPIRE as of 21 Oct 2026
	
	%\cite{Ringwald:2012hr}
	\bibitem{Ringwald:2012hr}
	A.~Ringwald,
	%``Exploring the Role of Axions and Other WISPs in the Dark Universe,''
	Phys. Dark Univ. \textbf{1}, 116-135 (2012)
	doi:10.1016/j.dark.2012.10.008
	[arXiv:1210.5081 [hep-ph]].
	%395 citations counted in INSPIRE as of 19 Oct 2026
	
	%\cite{Irastorza:2018dyq}
	\bibitem{Irastorza:2018dyq}
	I.~G.~Irastorza and J.~Redondo,
	%``New experimental approaches in the search for axion-like particles,''
	Prog. Part. Nucl. Phys. \textbf{102}, 89-159 (2018)
	doi:10.1016/j.ppnp.2018.05.003
	[arXiv:1801.08127 [hep-ph]].
	%1036 citations counted in INSPIRE as of 22 Oct 2026
	
	%\cite{DiLuzio:2020wdo}
	\bibitem{DiLuzio:2020wdo}
	L.~Di Luzio, M.~Giannotti, E.~Nardi and L.~Visinelli,
	%``The landscape of QCD axion models,''
	Phys. Rept. \textbf{870}, 1-117 (2020)
	doi:10.1016/j.physrep.2020.06.002
	[arXiv:2003.01100 [hep-ph]].
	%1220 citations counted in INSPIRE as of 22 Oct 2026
	
	%\cite{Mimasu:2014nea}
	\bibitem{Mimasu:2014nea}
	K.~Mimasu and V.~Sanz,
	%``ALPs at Colliders,''
	JHEP \textbf{06}, 173 (2015)
	doi:10.1007/JHEP06(2015)173
	[arXiv:1409.4792 [hep-ph]].
	%274 citations counted in INSPIRE as of 19 Oct 2026
	
	%\cite{Jaeckel:2015jla}
	\bibitem{Jaeckel:2015jla}
	J.~Jaeckel and M.~Spannowsky,
	%``Probing MeV to 90 GeV axion-like particles with LEP and LHC,''
	Phys. Lett. B \textbf{753}, 482-487 (2016)
	doi:10.1016/j.physletb.2015.12.037
	[arXiv:1509.00476 [hep-ph]].
	%385 citations counted in INSPIRE as of 21 Oct 2026
	
	%\cite{Bauer:2017ris}
	\bibitem{Bauer:2017ris}
	M.~Bauer, M.~Neubert and A.~Thamm,
	%``Collider Probes of Axion-Like Particles,''
	JHEP \textbf{12}, 044 (2017)
	doi:10.1007/JHEP12(2017)044
	[arXiv:1708.00443 [hep-ph]].
	%703 citations counted in INSPIRE as of 21 Oct 2026
	
	%\cite{Brivio:2017ije}
	\bibitem{Brivio:2017ije}
	I.~Brivio, M.~B.~Gavela, L.~Merlo, K.~Mimasu, J.~M.~No, R.~del Rey and V.~Sanz,
	%``ALPs Effective Field Theory and Collider Signatures,''
	Eur. Phys. J. C \textbf{77}, no.8, 572 (2017)
	doi:10.1140/epjc/s10052-017-5111-3
	[arXiv:1701.05379 [hep-ph]].
	%353 citations counted in INSPIRE as of 21 Oct 2026
	
	%\cite{Mariotti:2017vtv}
	\bibitem{Mariotti:2017vtv}
	A.~Mariotti, D.~Redigolo, F.~Sala and K.~Tobioka,
	%``New LHC bound on low-mass diphoton resonances,''
	Phys. Lett. B \textbf{783}, 13-18 (2018)
	doi:10.1016/j.physletb.2018.06.039
	[arXiv:1710.01743 [hep-ph]].
	%158 citations counted in INSPIRE as of 20 Oct 2026
	
	%\cite{Dolan:2017osp}
	\bibitem{Dolan:2017osp}
	M.~J.~Dolan, T.~Ferber, C.~Hearty, F.~Kahlhoefer and K.~Schmidt-Hoberg,
	%``Revised constraints and Belle II sensitivity for visible and invisible axion-like particles,''
	JHEP \textbf{12}, 094 (2017)
	[erratum: JHEP \textbf{03}, 190 (2021)]
	doi:10.1007/JHEP12(2017)094
	[arXiv:1709.00009 [hep-ph]].
	%366 citations counted in INSPIRE as of 20 Oct 2026
	
	%\cite{Alonso-Alvarez:2018irt}
	\bibitem{Alonso-Alvarez:2018irt}
	G.~Alonso-{\'A}lvarez, M.~B.~Gavela and P.~Quilez,
	%``Axion couplings to electroweak gauge bosons,''
	Eur. Phys. J. C \textbf{79}, no.3, 223 (2019)
	doi:10.1140/epjc/s10052-019-6732-5
	[arXiv:1811.05466 [hep-ph]].
	%112 citations counted in INSPIRE as of 20 Oct 2026
	
	%\cite{Gavela:2019wzg}
	\bibitem{Gavela:2019wzg}
	M.~B.~Gavela, R.~Houtz, P.~Quilez, R.~Del Rey and O.~Sumensari,
	%``Flavor constraints on electroweak ALP couplings,''
	Eur. Phys. J. C \textbf{79}, no.5, 369 (2019)
	doi:10.1140/epjc/s10052-019-6889-y
	[arXiv:1901.02031 [hep-ph]].
	%129 citations counted in INSPIRE as of 13 Oct 2026
	
	%\cite{Chala:2020wvs}
	\bibitem{Chala:2020wvs}
	M.~Chala, G.~Guedes, M.~Ramos and J.~Santiago,
	%``Running in the ALPs,''
	Eur. Phys. J. C \textbf{81}, no.2, 181 (2021)
	doi:10.1140/epjc/s10052-021-08968-2
	[arXiv:2012.09017 [hep-ph]].
	%157 citations counted in INSPIRE as of 13 Oct 2026
	
	%\cite{Bertulani:1987tz}
	\bibitem{Bertulani:1987tz}
	C.~A.~Bertulani and G.~Baur,
	%``Electromagnetic Processes in Relativistic Heavy Ion Collisions,''
	Phys. Rept. \textbf{163}, 299 (1988)
	doi:10.1016/0370-1573(88)90142-1
	%841 citations counted in INSPIRE as of 12 Oct 2026
	
	%\cite{Vidovic:1992ik}
	\bibitem{Vidovic:1992ik}
	M.~Vidovic, M.~Greiner, C.~Best and G.~Soff,
	%``Impact parameter dependence of the electromagnetic particle production in ultrarelativistic heavy ion collisions,''
	Phys. Rev. C \textbf{47}, 2308-2319 (1993)
	doi:10.1103/PhysRevC.47.2308
	%159 citations counted in INSPIRE as of 22 Oct 2026
	
	%\cite{Fermi:1924tc}
	\bibitem{Fermi:1924tc}
	E.~Fermi,
	%``On the Theory of the impact between atoms and electrically charged particles,''
	Z. Phys. \textbf{29}, 315-327 (1924)
	doi:10.1007/BF03184853
	%438 citations counted in INSPIRE as of 21 Oct 2026
	
	%\cite{vonWeizsacker:1934nji}
	\bibitem{vonWeizsacker:1934nji}
	C.~F.~von Weizsacker,
	%``Radiation emitted in collisions of very fast electrons,''
	Z. Phys. \textbf{88}, 612-625 (1934)
	doi:10.1007/BF01333110
	%1479 citations counted in INSPIRE as of 22 Oct 2026
	
	%\cite{Williams:1934ad}
	\bibitem{Williams:1934ad}
	E.~J.~Williams,
	%``Nature of the high-energy particles of penetrating radiation and status of ionization and radiation formulae,''
	Phys. Rev. \textbf{45}, 729-730 (1934)
	doi:10.1103/PhysRev.45.729
	%1143 citations counted in INSPIRE as of 22 Oct 2026
	
	%\cite{Budnev:1974de}
	\bibitem{Budnev:1974de}
	V.~M.~Budnev, I.~F.~Ginzburg, G.~V.~Meledin and V.~G.~Serbo,
	%``The Two photon particle production mechanism. Physical problems. Applications. Equivalent photon approximation,''
	Phys. Rept. \textbf{15}, 181-281 (1975)
	doi:10.1016/0370-1573(75)90009-5
	%1612 citations counted in INSPIRE as of 20 Oct 2026
	
	%\cite{Baur:2001jj}
	\bibitem{Baur:2001jj}
	G.~Baur, K.~Hencken, D.~Trautmann, S.~Sadovsky and Y.~Kharlov,
	%``Coherent gamma gamma and gamma-A interactions in very peripheral collisions at relativistic ion colliders,''
	Phys. Rept. \textbf{364}, 359-450 (2002)
	doi:10.1016/S0370-1573(01)00101-6
	[arXiv:hep-ph/0112211 [hep-ph]].
	%478 citations counted in INSPIRE as of 20 Oct 2026
	
	%\cite{Bertulani:2005ru}
	\bibitem{Bertulani:2005ru}
	C.~A.~Bertulani, S.~R.~Klein and J.~Nystrand,
	%``Physics of ultra-peripheral nuclear collisions,''
	Ann. Rev. Nucl. Part. Sci. \textbf{55}, 271-310 (2005)
	doi:10.1146/annurev.nucl.55.090704.151526
	[arXiv:nucl-ex/0502005 [nucl-ex]].
	%628 citations counted in INSPIRE as of 22 Oct 2026
	
	%\cite{Baltz:2007kq}
	\bibitem{Baltz:2007kq}
	A.~J.~Baltz, G.~Baur, D.~d'Enterria, L.~Frankfurt, F.~Gelis, V.~Guzey, K.~Hencken, Y.~Kharlov, M.~Klasen and S.~R.~Klein, \textit{et al.}
	%``The Physics of Ultraperipheral Collisions at the LHC,''
	Phys. Rept. \textbf{458}, 1-171 (2008)
	doi:10.1016/j.physrep.2007.12.001
	[arXiv:0706.3356 [nucl-ex]].
	%803 citations counted in INSPIRE as of 22 Oct 2026
	
	%\cite{Klein:2020fmr}
	\bibitem{Klein:2020fmr}
	S.~Klein and P.~Steinberg,
	%``Photonuclear and Two-photon Interactions at High-Energy Nuclear Colliders,''
	Ann. Rev. Nucl. Part. Sci. \textbf{70}, 323-354 (2020)
	doi:10.1146/annurev-nucl-030320-033923
	[arXiv:2005.01872 [nucl-ex]].
	%168 citations counted in INSPIRE as of 22 Oct 2026
	
	%\cite{STAR:2019wlg}
	\bibitem{STAR:2019wlg}
	J.~Adam \textit{et al.} [STAR],
	%``Measurement of $e^+e^-$ Momentum and Angular Distributions from Linearly Polarized Photon Collisions,''
	Phys. Rev. Lett. \textbf{127}, no.5, 052302 (2021)
	doi:10.1103/PhysRevLett.127.052302
	[arXiv:1910.12400 [nucl-ex]].
	%219 citations counted in INSPIRE as of 22 Oct 2026
	
	%\cite{ATLAS:2020epq}
	\bibitem{ATLAS:2020epq}
	G.~Aad \textit{et al.} [ATLAS],
	%``Exclusive dimuon production in ultraperipheral Pb+Pb collisions at $\sqrt{s_{\mathrm{NN}}} = 5.02$ TeV with ATLAS,''
	Phys. Rev. C \textbf{104}, 024906 (2021)
	doi:10.1103/PhysRevC.104.024906
	[arXiv:2011.12211 [nucl-ex]].
	%116 citations counted in INSPIRE as of 22 Oct 2026
	
	%\cite{ATLAS:2018pfw}
	\bibitem{ATLAS:2018pfw}
	M.~Aaboud \textit{et al.} [ATLAS],
	%``Observation of centrality-dependent acoplanarity for muon pairs produced via two-photon scattering in Pb+Pb collisions at $\sqrt{s_{\mathrm{NN}}}=5.02$ TeV with the ATLAS detector,''
	Phys. Rev. Lett. \textbf{121}, no.21, 212301 (2018)
	doi:10.1103/PhysRevLett.121.212301
	[arXiv:1806.08708 [nucl-ex]].
	%113 citations counted in INSPIRE as of 02 Oct 2026
	
	%\cite{ATLAS:2022yad}
	\bibitem{ATLAS:2022yad}
	G.~Aad \textit{et al.} [ATLAS],
	%``Measurement of muon pairs produced via {\ensuremath{\gamma}}{\ensuremath{\gamma}} scattering in nonultraperipheral Pb+Pb collisions at sNN=5.02 TeV with the ATLAS detector,''
	Phys. Rev. C \textbf{107}, no.5, 054907 (2023)
	doi:10.1103/PhysRevC.107.054907
	[arXiv:2206.12594 [nucl-ex]].
	%32 citations counted in INSPIRE as of 30 Sep 2026
	
	%\cite{Dai:2024imb}
	\bibitem{Dai:2024imb}
	J.~P.~Dai and S.~Zhao,
	%``Production of true para-muonium in linearly polarized photon fusions,''
	Phys. Rev. D \textbf{109}, no.5, 054022 (2024)
	doi:10.1103/PhysRevD.109.054022
	[arXiv:2401.04681 [hep-ph]].
	%10 citations counted in INSPIRE as of 05 Oct 2026
	
	%\cite{dEnterria:2025ecx}
	\bibitem{dEnterria:2025ecx}
	D.~d'Enterria and K.~Kang,
	%``Exclusive photon-fusion production of even-spin resonances and exotic QED atoms in high-energy hadron collisions,''
	Phys. Rev. D \textbf{112}, no.11, 116022 (2025)
	doi:10.1103/rnxl-v6gd
	[arXiv:2503.10952 [hep-ph]].
	%19 citations counted in INSPIRE as of 21 Oct 2026
	
	%\cite{Feng:2026yba}
	\bibitem{Feng:2026yba}
	Q.~M.~Feng, Q.~W.~Hu and C.~F.~Qiao,
	%``True Leptonium ($l^+ l^-$) Production in UPC Triphoton Interaction,''
	[arXiv:2604.21838 [hep-ph]].
	%2 citations counted in INSPIRE as of 26 Sep 2026
	
	%\cite{Knapen:2016moh}
	\bibitem{Knapen:2016moh}
	S.~Knapen, T.~Lin, H.~K.~Lou and T.~Melia,
	%``Searching for Axionlike Particles with Ultraperipheral Heavy-Ion Collisions,''
	Phys. Rev. Lett. \textbf{118}, no.17, 171801 (2017)
	doi:10.1103/PhysRevLett.118.171801
	[arXiv:1607.06083 [hep-ph]].
	%340 citations counted in INSPIRE as of 13 Oct 2026
	
	%\cite{CMS:2018erd}
	\bibitem{CMS:2018erd}
	A.~M.~Sirunyan \textit{et al.} [CMS],
	%``Evidence for light-by-light scattering and searches for axion-like particles in ultraperipheral PbPb collisions at $\sqrt{s_\mathrm{NN}} =$ 5.02 TeV,''
	Phys. Lett. B \textbf{797}, 134826 (2019)
	doi:10.1016/j.physletb.2019.134826
	[arXiv:1810.04602 [hep-ex]].
	%420 citations counted in INSPIRE as of 12 Oct 2026
	
	%\cite{ATLAS:2017fur}
	\bibitem{ATLAS:2017fur}
	M.~Aaboud \textit{et al.} [ATLAS],
	%``Evidence for light-by-light scattering in heavy-ion collisions with the ATLAS detector at the LHC,''
	Nature Phys. \textbf{13}, no.9, 852-858 (2017)
	doi:10.1038/nphys4208
	[arXiv:1702.01625 [hep-ex]].
	%513 citations counted in INSPIRE as of 16 Oct 2026
	
	%\cite{ATLAS:2020hii}
	\bibitem{ATLAS:2020hii}
	G.~Aad \textit{et al.} [ATLAS],
	%``Measurement of light-by-light scattering and search for axion-like particles with 2.2 nb$^{-1}$ of Pb+Pb data with the ATLAS detector,''
	JHEP \textbf{03}, 243 (2021)
	[erratum: JHEP \textbf{11}, 050 (2021)]
	doi:10.1007/JHEP03(2021)243
	[arXiv:2008.05355 [hep-ex]].
	%247 citations counted in INSPIRE as of 21 Oct 2026
	
	%\cite{Shao:2022cly}
	\bibitem{Shao:2022cly}
	H.~S.~Shao and D.~d'Enterria,
	%``gamma-UPC: automated generation of exclusive photon-photon processes in ultraperipheral proton and nuclear collisions with varying form factors,''
	JHEP \textbf{09}, 248 (2022)
	doi:10.1007/JHEP09(2022)248
	[arXiv:2207.03012 [hep-ph]].
	%103 citations counted in INSPIRE as of 21 Oct 2026
	
	%\cite{Kaidalov:2003fw}
	\bibitem{Kaidalov:2003fw}
	A.~B.~Kaidalov, V.~A.~Khoze, A.~D.~Martin and M.~G.~Ryskin,
	%``Central exclusive diffractive production as a spin-parity analyser: From Hadrons to Higgs,''
	Eur. Phys. J. C \textbf{31}, 387-396 (2003)
	doi:10.1140/epjc/s2003-01371-5
	[arXiv:hep-ph/0307064 [hep-ph]].
	%158 citations counted in INSPIRE as of 18 Aug 2026
	
	%\cite{Khoze:2004yb}
	\bibitem{Khoze:2004yb}
	V.~A.~Khoze, A.~D.~Martin, M.~G.~Ryskin and W.~J.~Stirling,
	%``Double diffractive chi meson production at the hadron colliders,''
	Eur. Phys. J. C \textbf{35}, 211-220 (2004)
	doi:10.1140/epjc/s2004-01857-6
	[arXiv:hep-ph/0403218 [hep-ph]].
	%102 citations counted in INSPIRE as of 07 Aug 2026
	
	%\cite{Harland-Lang:2016qjy}
	\bibitem{Harland-Lang:2016qjy}
	L.~A.~Harland-Lang, V.~A.~Khoze and M.~G.~Ryskin,
	%``The production of a diphoton resonance via photon-photon fusion,''
	JHEP \textbf{03}, 182 (2016)
	doi:10.1007/JHEP03(2016)182
	[arXiv:1601.07187 [hep-ph]].
	%78 citations counted in INSPIRE as of 12 Aug 2026
	
	%\cite{Belotsky:2004ex}
	\bibitem{Belotsky:2004ex}
	K.~Belotsky, V.~A.~Khoze, A.~D.~Martin and M.~G.~Ryskin,
	%``Can an invisible Higgs boson be seen via diffraction at the LHC?,''
	Eur. Phys. J. C \textbf{36}, 503-507 (2004)
	doi:10.1140/epjc/s2004-01967-1
	[arXiv:hep-ph/0406037 [hep-ph]].
	%34 citations counted in INSPIRE as of 04 Dec 2025
	
	%\cite{Khoze:2017igg}
	\bibitem{Khoze:2017igg}
	V.~A.~Khoze, A.~D.~Martin and M.~G.~Ryskin,
	%``Can invisible objects be {\textquoteleft}seen{\textquoteright} via forward proton detectors at the LHC?,''
	J. Phys. G \textbf{44}, no.5, 055002 (2017)
	doi:10.1088/1361-6471/aa6457
	[arXiv:1702.05023 [hep-ph]].
	%15 citations counted in INSPIRE as of 17 Jul 2026
	
	%\cite{DellAquila:1985mtb}
	\bibitem{DellAquila:1985mtb}
	J.~R.~Dell'Aquila and C.~A.~Nelson,
	%``$P$ or {CP} Determination by Sequential Decays: V1 V2 Modes With Decays Into $\bar{\ell}$epton (A) $\ell (B$) And/or $\bar{q}$ (A) $q (B$),''
	Phys. Rev. D \textbf{33}, 80 (1986)
	doi:10.1103/PhysRevD.33.80
	%101 citations counted in INSPIRE as of 14 Aug 2026
	
	%\cite{Plehn:2001nj}
	\bibitem{Plehn:2001nj}
	T.~Plehn, D.~L.~Rainwater and D.~Zeppenfeld,
	%``Determining the Structure of Higgs Couplings at the LHC,''
	Phys. Rev. Lett. \textbf{88}, 051801 (2002)
	doi:10.1103/PhysRevLett.88.051801
	[arXiv:hep-ph/0105325 [hep-ph]].
	%358 citations counted in INSPIRE as of 01 Oct 2026
	
	%\cite{Choi:2002jk}
	\bibitem{Choi:2002jk}
	S.~Y.~Choi, D.~J.~Miller, M.~M.~Muhlleitner and P.~M.~Zerwas,
	%``Identifying the Higgs spin and parity in decays to Z pairs,''
	Phys. Lett. B \textbf{553}, 61-71 (2003)
	doi:10.1016/S0370-2693(02)03191-X
	[arXiv:hep-ph/0210077 [hep-ph]].
	%280 citations counted in INSPIRE as of 30 Sep 2026
	
	%\cite{Buszello:2002uu}
	\bibitem{Buszello:2002uu}
	C.~P.~Buszello, I.~Fleck, P.~Marquard and J.~J.~van der Bij,
	%``Prospective analysis of spin- and CP-sensitive variables in H ---{\ensuremath{>}} Z Z ---{\ensuremath{>}} l(1)+ l(1)- l(2)+ l(2)- at the LHC,''
	Eur. Phys. J. C \textbf{32}, 209-219 (2004)
	doi:10.1140/epjc/s2003-01392-0
	[arXiv:hep-ph/0212396 [hep-ph]].
	%209 citations counted in INSPIRE as of 16 Sep 2026
	
	%\cite{Berge:2008wi}
	\bibitem{Berge:2008wi}
	S.~Berge, W.~Bernreuther and J.~Ziethe,
	%``Determining the CP parity of Higgs bosons at the LHC in their tau decay channels,''
	Phys. Rev. Lett. \textbf{100}, 171605 (2008)
	doi:10.1103/PhysRevLett.100.171605
	[arXiv:0801.2297 [hep-ph]].
	%90 citations counted in INSPIRE as of 30 Sep 2026
	
	%\cite{Gao:2010qx}
	\bibitem{Gao:2010qx}
	Y.~Gao, A.~V.~Gritsan, Z.~Guo, K.~Melnikov, M.~Schulze and N.~V.~Tran,
	%``Spin Determination of Single-Produced Resonances at Hadron Colliders,''
	Phys. Rev. D \textbf{81}, 075022 (2010)
	doi:10.1103/PhysRevD.81.075022
	[arXiv:1001.3396 [hep-ph]].
	%566 citations counted in INSPIRE as of 13 Oct 2026
	
	%\cite{Bolognesi:2012mm}
	\bibitem{Bolognesi:2012mm}
	S.~Bolognesi, Y.~Gao, A.~V.~Gritsan, K.~Melnikov, M.~Schulze, N.~V.~Tran and A.~Whitbeck,
	%``On the Spin and Parity of a Single-Produced Resonance at the LHC,''
	Phys. Rev. D \textbf{86}, 095031 (2012)
	doi:10.1103/PhysRevD.86.095031
	[arXiv:1208.4018 [hep-ph]].
	%502 citations counted in INSPIRE as of 21 Oct 2026
	
	%\cite{Anderson:2013afp}
	\bibitem{Anderson:2013afp}
	I.~Anderson, S.~Bolognesi, F.~Caola, Y.~Gao, A.~V.~Gritsan, C.~B.~Martin, K.~Melnikov, M.~Schulze, N.~V.~Tran and A.~Whitbeck, \textit{et al.}
	%``Constraining Anomalous HVV Interactions at Proton and Lepton Colliders,''
	Phys. Rev. D \textbf{89}, no.3, 035007 (2014)
	doi:10.1103/PhysRevD.89.035007
	[arXiv:1309.4819 [hep-ph]].
	%286 citations counted in INSPIRE as of 01 Oct 2026
	
	%\cite{Ellis:2012jv}
	\bibitem{Ellis:2012jv}
	J.~Ellis, R.~Fok, D.~S.~Hwang, V.~Sanz and T.~You,
	%``Distinguishing 'Higgs' spin hypotheses using $\gamma \gamma$ and $W W^*$ decays,''
	Eur. Phys. J. C \textbf{73}, 2488 (2013)
	doi:10.1140/epjc/s10052-013-2488-5
	[arXiv:1210.5229 [hep-ph]].
	%72 citations counted in INSPIRE as of 07 Aug 2026
	
	%\cite{Freitas:2012kw}
	\bibitem{Freitas:2012kw}
	A.~Freitas and P.~Schwaller,
	%``Higgs CP Properties From Early LHC Data,''
	Phys. Rev. D \textbf{87}, no.5, 055014 (2013)
	doi:10.1103/PhysRevD.87.055014
	[arXiv:1211.1980 [hep-ph]].
	%66 citations counted in INSPIRE as of 16 Sep 2026
	
	%\cite{Harnik:2013aja}
	\bibitem{Harnik:2013aja}
	R.~Harnik, A.~Martin, T.~Okui, R.~Primulando and F.~Yu,
	%``Measuring CP Violation in $h \to \tau^+ \tau^-$ at Colliders,''
	Phys. Rev. D \textbf{88}, no.7, 076009 (2013)
	doi:10.1103/PhysRevD.88.076009
	[arXiv:1308.1094 [hep-ph]].
	%112 citations counted in INSPIRE as of 16 Oct 2026
	
	%\cite{Drees:1988pp}
	\bibitem{Drees:1988pp}
	M.~Drees and D.~Zeppenfeld,
	%``Production of Supersymmetric Particles in Elastic $e p$ Collisions,''
	Phys. Rev. D \textbf{39}, 2536 (1989)
	doi:10.1103/PhysRevD.39.2536
	%190 citations counted in INSPIRE as of 20 Oct 2026
	
	%\cite{Kniehl:1990iv}
	\bibitem{Kniehl:1990iv}
	B.~A.~Kniehl,
	%``Elastic e p scattering and the Weizsacker-Williams approximation,''
	Phys. Lett. B \textbf{254}, 267-273 (1991)
	doi:10.1016/0370-2693(91)90432-P
	%128 citations counted in INSPIRE as of 06 Aug 2026
	
	%\cite{Albrow:2008pn}
	\bibitem{Albrow:2008pn}
	M.~G.~Albrow \textit{et al.} [FP420 R{\&}D],
	%``The FP420 {\textbackslash}{\&}  Project: Higgs and New Physics with forward protons at the LHC,''
	JINST \textbf{4}, T10001 (2009)
	doi:10.1088/1748-0221/4/10/T10001
	[arXiv:0806.0302 [hep-ex]].
	%303 citations counted in INSPIRE as of 20 Oct 2026
	
	%\cite{Chapon:2009hh}
	\bibitem{Chapon:2009hh}
	E.~Chapon, C.~Royon and O.~Kepka,
	%``Anomalous quartic W W gamma gamma, Z Z gamma gamma, and trilinear WW gamma couplings in two-photon processes at high luminosity at the LHC,''
	Phys. Rev. D \textbf{81}, 074003 (2010)
	doi:10.1103/PhysRevD.81.074003
	[arXiv:0912.5161 [hep-ph]].
	%232 citations counted in INSPIRE as of 11 Oct 2026
	
	%\cite{Fichet:2013gsa}
	\bibitem{Fichet:2013gsa}
	S.~Fichet, G.~von Gersdorff, O.~Kepka, B.~Lenzi, C.~Royon and M.~Saimpert,
	%``Probing new physics in diphoton production with proton tagging at the Large Hadron Collider,''
	Phys. Rev. D \textbf{89}, 114004 (2014)
	doi:10.1103/PhysRevD.89.114004
	[arXiv:1312.5153 [hep-ph]].
	%118 citations counted in INSPIRE as of 11 Oct 2026
	
	%\cite{Fichet:2014uka}
	\bibitem{Fichet:2014uka}
	S.~Fichet, G.~von Gersdorff, B.~Lenzi, C.~Royon and M.~Saimpert,
	%``Light-by-light scattering with intact protons at the LHC: from Standard Model to New Physics,''
	JHEP \textbf{02}, 165 (2015)
	doi:10.1007/JHEP02(2015)165
	[arXiv:1411.6629 [hep-ph]].
	%154 citations counted in INSPIRE as of 16 Sep 2026
	
	%\cite{Fichet:2015vvy}
	\bibitem{Fichet:2015vvy}
	S.~Fichet, G.~von Gersdorff and C.~Royon,
	%``Scattering light by light at 750 GeV at the LHC,''
	Phys. Rev. D \textbf{93}, no.7, 075031 (2016)
	doi:10.1103/PhysRevD.93.075031
	[arXiv:1512.05751 [hep-ph]].
	%243 citations counted in INSPIRE as of 06 Aug 2026
	
	%\cite{Harland-Lang:2016apc}
	\bibitem{Harland-Lang:2016apc}
	L.~A.~Harland-Lang, V.~A.~Khoze and M.~G.~Ryskin,
	%``The photon PDF in events with rapidity gaps,''
	Eur. Phys. J. C \textbf{76}, no.5, 255 (2016)
	doi:10.1140/epjc/s10052-016-4100-2
	[arXiv:1601.03772 [hep-ph]].
	%91 citations counted in INSPIRE as of 20 Oct 2026
	
	%\cite{Baldenegro:2018hng}
	\bibitem{Baldenegro:2018hng}
	C.~Baldenegro, S.~Fichet, G.~von Gersdorff and C.~Royon,
	%``Searching for axion-like particles with proton tagging at the LHC,''
	JHEP \textbf{06}, 131 (2018)
	doi:10.1007/JHEP06(2018)131
	[arXiv:1803.10835 [hep-ph]].
	%134 citations counted in INSPIRE as of 09 Sep 2026
	
	%\cite{CMS:2022hly}
	\bibitem{CMS:2022hly}
	A.~Tumasyan \textit{et al.} [CMS and TOTEM],
	%``Proton reconstruction with the CMS-TOTEM Precision Proton Spectrometer,''
	JINST \textbf{18}, no.09, P09009 (2023)
	doi:10.1088/1748-0221/18/09/P09009
	[arXiv:2210.05854 [hep-ex]].
	%29 citations counted in INSPIRE as of 04 Oct 2026
	
	%\cite{ATLAS:2023zfc}
	\bibitem{ATLAS:2023zfc}
	G.~Aad \textit{et al.} [ATLAS],
	%``Search for an axion-like particle with forward proton scattering in association with photon pairs at ATLAS,''
	JHEP \textbf{07}, 234 (2023)
	doi:10.1007/JHEP07(2023)234
	[arXiv:2304.10953 [hep-ex]].
	%52 citations counted in INSPIRE as of 21 Oct 2026
	
	%\cite{CMS:2023jgd}
	\bibitem{CMS:2023jgd}
	A.~Tumasyan \textit{et al.} [TOTEM and CMS],
	%``Search for high-mass exclusive diphoton production with tagged protons in proton-proton collisions at s=13{\,}{\,}TeV,''
	Phys. Rev. D \textbf{110}, no.1, 012010 (2024)
	doi:10.1103/PhysRevD.110.012010
	[arXiv:2311.02725 [hep-ex]].
	%41 citations counted in INSPIRE as of 21 Oct 2026
	
	%\cite{Mulders:2000sh}
	\bibitem{Mulders:2000sh}
	P.~J.~Mulders and J.~Rodrigues,
	%``Transverse momentum dependence in gluon distribution and fragmentation functions,''
	Phys. Rev. D \textbf{63}, 094021 (2001)
	doi:10.1103/PhysRevD.63.094021
	[arXiv:hep-ph/0009343 [hep-ph]].
	%425 citations counted in INSPIRE as of 22 Oct 2026
	
	%\cite{Boer:2011kf}
	\bibitem{Boer:2011kf}
	D.~Boer, W.~J.~den Dunnen, C.~Pisano, M.~Schlegel and W.~Vogelsang,
	%``Linearly Polarized Gluons and the Higgs Transverse Momentum Distribution,''
	Phys. Rev. Lett. \textbf{108}, 032002 (2012)
	doi:10.1103/PhysRevLett.108.032002
	[arXiv:1109.1444 [hep-ph]].
	%142 citations counted in INSPIRE as of 20 Oct 2026
	
	%\cite{Harland-Lang:2020veo}
	\bibitem{Harland-Lang:2020veo}
	L.~A.~Harland-Lang, M.~Tasevsky, V.~A.~Khoze and M.~G.~Ryskin,
	%``A new approach to modelling elastic and inelastic photon-initiated production at the LHC: SuperChic 4,''
	Eur. Phys. J. C \textbf{80}, no.10, 925 (2020)
	doi:10.1140/epjc/s10052-020-08455-0
	[arXiv:2007.12704 [hep-ph]].
	%107 citations counted in INSPIRE as of 22 Oct 2026
	
	%\cite{Khoze:2013dha}
	\bibitem{Khoze:2013dha}
	V.~A.~Khoze, A.~D.~Martin and M.~G.~Ryskin,
	%``Diffraction at the LHC,''
	Eur. Phys. J. C \textbf{73}, 2503 (2013)
	doi:10.1140/epjc/s10052-013-2503-x
	[arXiv:1306.2149 [hep-ph]].
	%85 citations counted in INSPIRE as of 15 Sep 2026
	
	%\cite{TOTEM:2023ewz}
	\bibitem{TOTEM:2023ewz}
	A.~Tumasyan \textit{et al.} [TOTEM and CMS],
	%``Search for high-mass exclusive diphoton production with tagged protons in proton-proton collisions at s=13{\,}{\,}TeV,''
	Phys. Rev. D \textbf{110}, no.1, 012010 (2024)
	doi:10.1103/PhysRevD.110.012010
	[arXiv:2311.02725 [hep-ex]].
	%41 citations counted in INSPIRE as of 21 Oct 2026
	
	%\cite{Cerny:2020rvp}
	\bibitem{Cerny:2020rvp}
	K.~{\v{C}}ern{\'y}, T.~S{\'y}kora, M.~Ta{\v{s}}evsk{\'y} and R.~{\v{Z}}leb{\v{c}}{\'\i}k,
	%``Performance studies of Time-of-Flight detectors at LHC,''
	JINST \textbf{16}, no.01, P01030 (2021)
	doi:10.1088/1748-0221/16/01/P01030
	[arXiv:2010.00237 [hep-ph]].
	%21 citations counted in INSPIRE as of 24 Sep 2026
	
	%\cite{Harland-Lang:2018hmi}
	\bibitem{Harland-Lang:2018hmi}
	L.~A.~Harland-Lang, V.~A.~Khoze, M.~G.~Ryskin and M.~Tasevsky,
	%``LHC Searches for Dark Matter in Compressed Mass Scenarios: Challenges in the Forward Proton Mode,''
	JHEP \textbf{04}, 010 (2019)
	doi:10.1007/JHEP04(2019)010
	[arXiv:1812.04886 [hep-ph]].
	%59 citations counted in INSPIRE as of 06 Oct 2026
	
	%\cite{Goncalves:2020saa}
	\bibitem{Goncalves:2020saa}
	V.~P.~Gon{\c{c}}alves, D.~E.~Martins, M.~S.~Rangel and M.~Tasevsky,
	%``Top quark pair production in the exclusive processes at the LHC,''
	Phys. Rev. D \textbf{102}, no.7, 074014 (2020)
	doi:10.1103/PhysRevD.102.074014
	[arXiv:2007.04565 [hep-ph]].
	%34 citations counted in INSPIRE as of 10 Sep 2026
	
	%\cite{Martins:2022dfg}
	\bibitem{Martins:2022dfg}
	D.~E.~Martins, M.~Tasevsky and V.~P.~Goncalves,
	%``Challenging exclusive top quark pair production at low and high luminosity LHC,''
	Phys. Rev. D \textbf{105}, no.11, 114002 (2022)
	doi:10.1103/PhysRevD.105.114002
	[arXiv:2202.01257 [hep-ph]].
	%23 citations counted in INSPIRE as of 06 Oct 2026
	
	%\cite{Tasevsky:2022sch}
	\bibitem{Tasevsky:2022sch}
	M.~Tasevsky, V.~Khoze, D.~Milne and M.~Ryskin,
	%``Searches for QCD instantons with forward proton tagging,''
	Eur. Phys. J. C \textbf{83}, no.1, 35 (2023)
	doi:10.1140/epjc/s10052-023-11178-7
	[arXiv:2208.14089 [hep-ph]].
	%9 citations counted in INSPIRE as of 06 Oct 2026

\end{thebibliography}
\end{document}